\documentclass[11pt]{article}
\usepackage{fa2025}
\usepackage{amsmath}
\usepackage{cite}
\usepackage{url}
\usepackage{graphicx}
\usepackage{color}
\usepackage{siunitx}
\usepackage[utf8]{inputenc}
\usepackage{subcaption}
\usepackage{enumitem}

\usepackage[acronym]{glossaries}
\usepackage{textcomp}
\newacronym{sma}{SMA}{Spherical Microphone Array}
\newacronym{hom}{HOM}{Higher Order Microphone}
\newacronym{ula}{ULA}{Uniform Linear Array}
\newacronym{toa}{TOA}{Time of Arrival}
\newacronym{rir}{RIR}{Room Impulse Response}
\newacronym{arir}{A-RIR}{Ambisonic Room Impulse Response}
\newacronym{mls}{MLS} {Maximum Length Sequence}
\newacronym{irs}{IRS} {Inverse Repeated Sequence}
\newacronym{vr-ptolemaic}{VR-PTOLEMAIC} {Virtual Reality Perceptual Testing Of Listening Environments Modeled through A-RIRs Interactive Convolution}
\newacronym{mushra}{MUSHRA}{MUlti-Stimulus test with Hidden Reference and Anchor}
\newacronym{gui}{GUI}{Graphical User Interface}
\newacronym{ui}{UI}{User Interface}
\newacronym{ux}{UX}{User Experience}
\newacronym{vr}{VR}{Virtual Reality}
\newacronym{ar}{AR}{Augmented Reality}
\newacronym{udp}{UDP}{User Datagram Protocol}
\newacronym{osc}{OSC}{Open Sound Control}
\newacronym{6dof}{6DoF}{6 Degrees of Freedom}
\newacronym{ve}{VE}{Virtual Environment}
\newacronym{pc}{PC}{Personal Computer}
\newacronym{3d}{3D}{Three Dimensional}
\newacronym{admm}{ADMM}{Alternating Direction Method of Multipliers}
\newacronym{sfr}{SFR}{Sound Field Reconstruction}
\newacronym{itu}{ITU}{International Telecommunications Union}
\newacronym{hmd}{HMD}{Head-Mounted Display}
\newacronym{hrtf}{HRTF}{Head-Related Transfer Function}

\title{VR-PTOLEMAIC: A Virtual Environment for the Perceptual Testing of Spatial Audio Algorithms}
\multauthor
{Paolo Ostan$^{*}$ \hspace{1cm} Francesca Del Gaudio \hspace{1cm} Federico Miotello} { \bfseries{Mirco Pezzoli \hspace{1cm} Fabio Antonacci}\\
  \textit{Dipartimento di Elettronica, Informazione e Bioingegneria - Politecnico di Milano}\\
Via Ponzio 34/5, 20133 Milano, Italia\\
\correspondingauthor{paolo.ostan@polimi.it}{Paolo Ostan}
}

\begin{document}
\maketitle
\begin{abstract}
%The perceptual evaluation of sound field reconstruction algorithms is an important step in the development of immersive audio applications.
%It ensures that synthesized sound fields meet quality standards in terms of listening experience, accounting for spatial perception and auditory realism.
%To support these evaluations, virtual reality can offer a powerful platform by providing immersive and interactive testing environments.
%In this paper, we present VR-PTOLEMAIC (Virtual Reality Perceptual Testing Of Listening Environments Modelled through Ambisonic-RIRs Interactive Convolution), a virtual reality evaluation system designed for assessing spatial audio algorithms.
%The system includes an implementation of the MUSHRA (MUlti-Stimulus test with Hidden Reference and Anchor) evaluation methodology, integrated into a virtual environment.
%In particular, users can position themselves in each of the 25 simulated listening positions of a virtually recreated seminar room (the very same featured in the HOMULA-RIR dataset) and evaluate simulated acoustic responses with respect to the actually recorded second-order ambisonic room impulse responses, all convolved with various source signals.
%Moreover, given the importance of human interaction in assessing spatial audio algorithms, the system allows users to dynamically rotate their listening position by means of a head-tracking device, providing a dynamic and realistic evaluation of sound field reconstruction algorithms.
The perceptual evaluation of spatial audio algorithms is an important step in the development of immersive audio applications, as it ensures that synthesized sound fields meet quality standards in terms of listening experience, spatial perception and auditory realism.
To support these evaluations, virtual reality can offer a powerful platform by providing immersive and interactive testing environments.
In this paper, we present VR-PTOLEMAIC, a virtual reality evaluation system designed for assessing spatial audio algorithms.
The system implements the MUSHRA (MUlti-Stimulus test with Hidden Reference and Anchor) evaluation methodology into a virtual environment.
In particular, users can position themselves in each of the 25 simulated listening positions of a virtually recreated seminar room and evaluate simulated acoustic responses with respect to the actually recorded second-order ambisonic room impulse responses, all convolved with various source signals.
We evaluated the usability of the proposed framework through an extensive testing campaign in which assessors were asked to compare the reconstruction capabilities of various sound field reconstruction algorithms.
Results show that the VR platform effectively supports the assessment of spatial audio algorithms, with generally positive feedback on user experience and immersivity.

\end{abstract}
\section{Introduction}\label{sec:introduction}
Spatial audio is a trending research field focused on understanding, recreating, and optimizing 3D auditory environments \cite{bassuet2014computational},
%By capturing or synthesizing the positional characteristics of sound sources and accurately rendering cues like direction, distance, and movement, spatial audio is able to
providing listeners with an immersive and interactive listening experience.
Applications range from virtual and augmented reality \cite{yang2022audio}  to gaming \cite{firat20223d}, teleconferencing and remote concerts \cite{patynen2011virtual}, where precise sound localization and an enhanced sense of presence enrich user engagement and perception.

The usual spatial audio pipeline \cite{rafaely2022spatial, cobos_overview_2022} begins with capturing the sound scene, followed by a processing stage where the spatial information is either modified or inferred (especially for aspects that cannot be measured directly) and ends with a reproduction phase in which the (potentially altered) scene is auralized.
Various specialized techniques each contribute to different aspects of this general pipeline.
Among the others, we can cite for example: microphone array processing \cite{politis2016microphone}, which leverages strategically arranged microphones to capture audio with higher spatial resolution; \acrfull{sfr} \cite{9632746, pezzoli_parametric_2020, miotello_reconstruction_2024, damiano2025zero, olivieri_physics-informed_2024, fernandez2023generative}, to recover the entire acoustic environment from limited microphone signals; directional source analysis \cite{pezzoli2022comparative}, which studies the spatial energy distribution of sound sources; \acrfull{hrtf} personalization \cite{sanchez2025towards}, that adapts the experience to an individual’s head and ear shape for a more accurate listening experience; and binaural reproduction methods \cite{ben2018joint} to deliver localized sound directly to each ear, mimicking natural listening and enhancing the sense of presence.
Although objective metrics play a critical role in objectively assessing many of such techniques, they might not fully capture the details that contribute to a listener’s sense of realism.
In fact, in spatial audio, the goal extends beyond merely reproducing an accurate acoustic field; it also involves rendering a plausible and immersive experience, where listeners feel as if they are part of the scene.
This subjective dimension of perception is inherently complex.
Consequently, robust evaluation of spatial audio systems relies not only on objective measurement \cite{ryu2008subjective} but also on subjective experiments that address how listeners actually experience the audio, assessing perceptual attributes such as localization accuracy, spatial clarity, immersion, and externalization.
%Designing, conducting, and interpreting these subjective evaluations demands careful consideration of experimental procedures—such as choosing representative stimuli, selecting test environments, and analyzing listener feedback—making the assessment of perceptual realism in spatial audio both indispensable and challenging.
%However, achieving high-quality spatial audio reproduction requires rigorous evaluation to ensure both technical accuracy and perceptual realism.
%The assessment of spatial audio audio systems relies on a combination of objective and subjective methodologies. Objective approaches quantify technical performance using physical measurements—such as frequency response, reverberation time, and early decay time \cite{ryu2008subjective}. While these metrics provide valuable insights, they do not fully capture the complex psychoacoustic phenomena that shape human perception of spatial sound \cite{guastavino2004perceptual}. Therefore, subjective evaluation is essential to assess perceptual attributes like localization accuracy, spatial clarity, immersion, and externalization.

%Subjective evaluation methods address this limitation by assessing perceptual attributes such as localization accuracy, spatial clarity, immersion, and externalization.
The standard approach for subjective sound quality evaluation is based on direct comparison methods, as outlined in the recommendations of the \acrfull{itu} \cite{itu2015bs, itu20151534, itu1996itu}. 
In recent years, methodologies such as the well-known \acrfull{mushra} \cite{series2014method} have become widely adopted, offering a structured framework for assessing quality variations between multiple stimuli relative to a reference. This approach enables participants to systematically compare different versions of an audio signal, facilitating a detailed evaluation of perceived differences across specific attributes.
Beyond direct comparison, other evaluation methodologies provide complementary insights into listeners' perception and evaluation of sound quality. These include sensory analysis protocols \cite{zacharov2018sensory}, indirect comparison methods \cite{wickelmaier2012scaling}, and behavioural analysis techniques \cite{pike2017direct}.
%During the years, with the rising importance of spatial audio these different evaluation techniques have been applied not only for audio quality evaluation but also for rating results in tasks like \acrshort{sfr}. This task represents an important for applications such as virtual acoustic reconstruction and acoustic scene navigation \cite{tylka2020fundamentals}. In particular, \acrshort{sfr} involves synthesizing an acoustic field over a target region using measurements from multiple microphones or estimated sound pressure and particle velocity data. However, challenges such as limited measurement data, environmental noise, and computational constraints affect performance. To address these issues, researchers employ different techniques like parametric models \cite{pezzoli_parametric_2020}, non-parametric models \cite{koyama_sparsity-based_2020} and data-driven machine learning approaches \cite{cobos_overview_2022}.

The need of evaluation techniques for spatial audio algorithms has led to adaptations of traditional subjective testing methodologies, resulting in the development of specialized test setups.
Free-field listening tests in controlled environments \cite{freigang_free-field_2014} and binaural playback evaluations using \acrshort{hrtf}s \cite{andreopoulou2015use} provide valuable insights into key spatial audio attributes such as localization accuracy and spatial perception.
Moreover, in the last few years, with the rise of \acrfull{vr} and \acrfull{ar}, spatial audio evaluation has gained new significance.
VR-based assessment methods \cite{medina2024assessing} offer a controlled yet immersive environment, enabling real-time testing of spatial audio reproduction in interactive scenarios. Thanks to the introduction of a visual experience, these methods enhance the perceptual realism of the virtual test scene and allow for precise evaluation of spatial immersion under dynamic conditions, supporting the refinement of spatial audio rendering techniques. Additionally, the ability to continuously track user behaviours, such as its movement and head rotation, provides valuable data for result analysis \cite{rummukainen2018audio}, offering deeper insights into how listeners interact with spatialized audio.
Finally, VR systems also enable the implementation of diverse test modalities, further expanding the scope of spatial audio evaluation \cite{gorzynski2020flexible}.

In this article, we introduce VR-PTOLEMAIC: a \acrfull{ve} for the perceptual testing of spatial audio algorithms.
Our framework integrates a real-world acoustically sampled environment (namely, a seminar room \cite{miotello_homula-rir_2024}) into a VR-based evaluation system.
Building on previous work on sound field synthesis for \acrfull{6dof} navigation \cite{rummukainen2018audio, mccormack_object-based_2022, figueroa2025reconstruction}, we render spatial audio using multiple second-order Ambisonic receivers, which listeners experience through head-tracked headphones while visually exploring the \acrshort{ve} via a VR headset.
By embedding the acoustical characteristics of a real room into a virtual setting, VR-PTOLEMAIC allows for direct comparison between measured and synthesized sound fields, enhancing the validity of subjective evaluations by exploiting a MUSHRA-like testing framework.
Moreover, in addition to traditional perceptual assessments, the system also incorporates real-time behavioural tracking, providing insights into how users interact with spatialized audio and the experience. This multimodal approach offers a tool for precisely evaluating spatial audio perception under realistic conditions.
To validate the proposed framework, we conducted an extensive evaluation campaign in which $15$ participants assessed various \acrshort{sfr} algorithms designed to reproduce the pressure field in the virtual seminar room.
The rest of the manuscript presents the evaluation system (Sec.~\ref{sec:ev_sys_arc}), its validation (Sec.~\ref{sec:eval_tests}) and a discussion on the results (Sec.~\ref{sec:discussion}).
\section{Evaluation System }\label{sec:ev_sys_arc}
\begin{figure}[tbp]
 \centerline{
 \includegraphics[width=7.5cm]{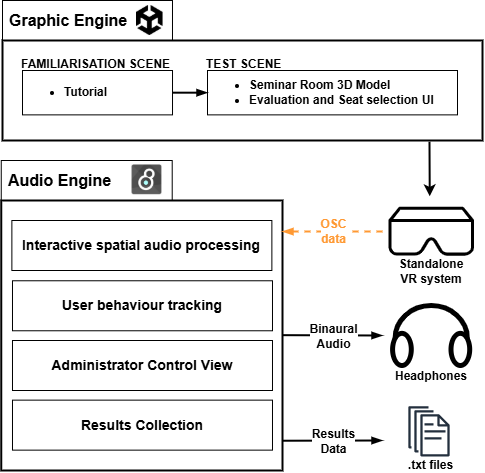}}
 \caption{Overview of the proposed evaluation platform.}
 \label{fig:implementation_scheme}
 \vspace{-2mm}
\end{figure}

    Perceptual tests for spatial audio in \acrshort{vr} require a multimodal environment that involves visual content, audio content and interactivity in order to maintain in the assessor a sense of presence and immersivity during the evaluation procedure.
    The following section presents a VR-based evaluation system for assessing spatial audio perception.
As shown in \figref{fig:implementation_scheme}, the platform integrates a VR application implemented in Unity\footnote{https://unity.com} and a Max\footnote{https://cycling74.com}-based audio processor, communicating via \acrfull{osc} over \acrfull{udp} for real-time interaction.
By means of a \acrshort{vr} headset with its controllers and headphones, users are able to navigate a \acrshort{3d} model of the test environment, select positions, and evaluate spatial audio attributes interacting with a \acrshort{mushra}-based \acrshort{gui}. The audio is processed dynamically allowing to listen and compare different \acrshort{sfr}s. The stimuli proposed are the response of a convolution between anechoic sound samples and measured or reconstructed \acrfullpl{arir}s.
Given that the visual content is rendered by the \acrshort{vr} headset itself, the platform does not require a \acrshort{vr}-ready \acrfull{pc} in order to perform test sessions.
Also, the software is designed for the Meta\footnote{https://www.meta.com} Quest 2 VR-headset, which allows for \acrshort{6dof} movement tracking without the need of external sensors, and the binaural audio reproduced on closed-ear headphones does not require specific characteristics for the test room.

\subsection{Visual content management}

    The first component of the platform is an application developed in Unity and preloaded on an all-in-one \acrshort{vr} system. It is used to immerse the assessor in a \acrshort{ve}, to interact with the \acrfull{gui} developed to perform the evaluations and to track user position and rotation.
    
\begin{figure}[tbp]
 \centerline{
 \includegraphics[width=7.8cm]{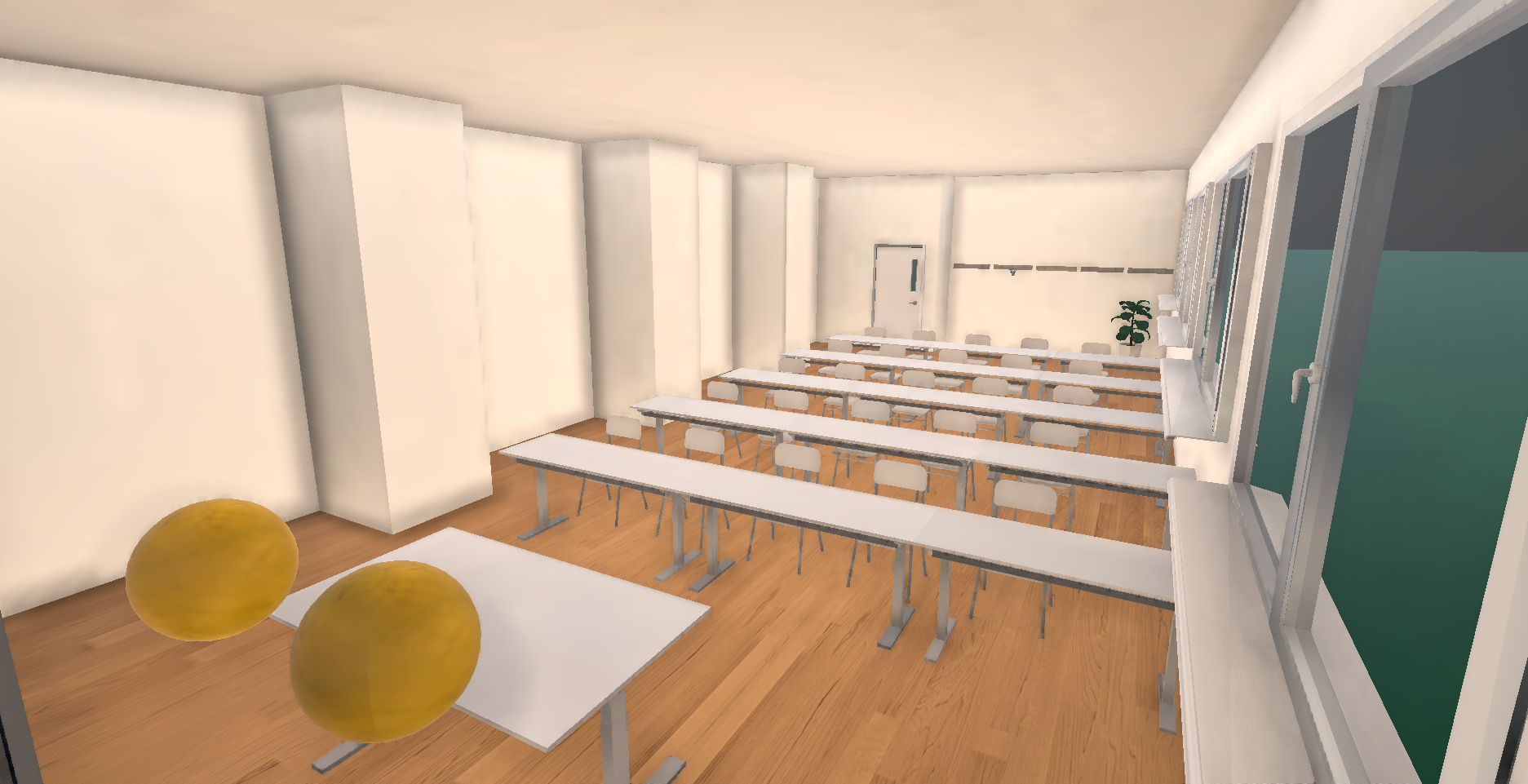}}
 \caption{The \acrshort{3d} reconstruction of the seminar room in \acrshort{vr}.}
 \label{fig:seminarroom_VR}
 \vspace{-2mm}
\end{figure}

\begin{figure}[ht]
\centering
    \begin{minipage}{6.8cm}
    \centering
        \begin{subfigure}[b]{\textwidth}
            \centering
            \includegraphics[width=\textwidth]{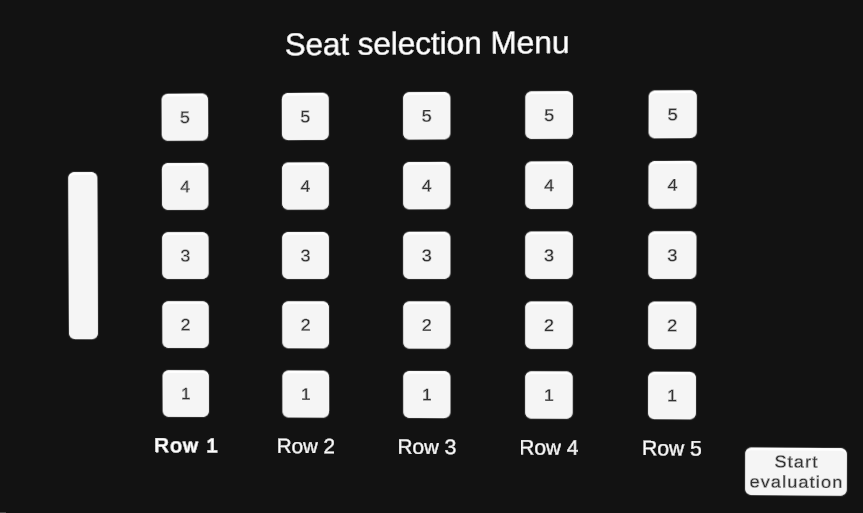}
            \caption{}
            \label{fig:seatselection_VR}
        \end{subfigure}
        \begin{subfigure}[b]{\textwidth}
            \centering
            \includegraphics[width=\textwidth]{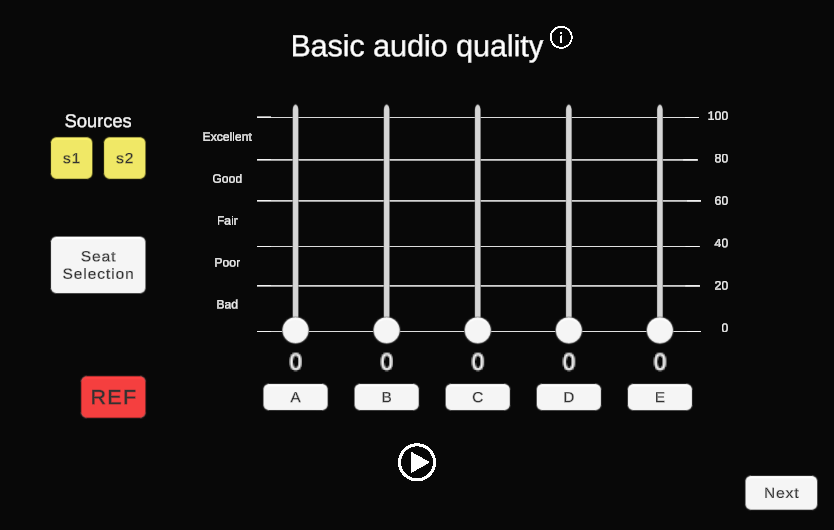}
            \caption{}
            \label{fig:MUSHRA_VR}
        \end{subfigure}
    \end{minipage}
    \caption{Seat selection \acrshort{ui} (a) and Virtual reality implementation of the \acrshort{mushra} \acrshort{ui} (b)}
    \label{fig:eval_ui}
    \vspace{-2mm}
\end{figure}

The \acrshort{ve} presented to the assessors (\figref{fig:seminarroom_VR}) is the \acrshort{3d} model of the seminar room presented in the HOMULA-RIR dataset \cite{miotello_homula-rir_2024}.
It can be noticed that, behind the main desk, 2 spheres are added to identify the location of the sound sources employed during the \acrshort{arir}s measurement campaign.
%, shown in \figref{fig:seminarroom_top} and \figref{fig:seminarroom}, in which the 2 spheres and the 25 chairs added, respectively identify the location of the sound sources and of the \acrlong{hom}s employed during the \acrshort{arir}s measurement campaign.
The system allows for potential free navigation with \acrshort{6dof}. However, due to the specific nature of the dataset \cite{miotello_homula-rir_2024}, in which the sound field is discretized into 25 spatial positions (corresponding to the position of the chairs in the room), user movement is constrained to these predefined points.
Throughout the evaluation session, the assessor can take full advantage of head rotation and exploit teleportation to move in the room by means of a dedicated \acrshort{ui} screen, represented in \figref{fig:seatselection_VR}.
The \acrshort{gui} for the evaluations (shown in \figref{fig:MUSHRA_VR}) is an implementation of the one presented in \cite{series2014method}, with the addition of a section for source selection and a button that opens the seat selection screen. In addition, an info button located next to the attribute title shows a short description of the attribute when clicked.
The \acrshort{ui} is attached to the left controller following its movement, so it can be moved in a convenient position by the user or activated and deactivated by means of a button on the same controller.

% \begin{figure}[ht]
% \centering
%     \begin{minipage}{6.8cm}
%     \centering
%         \begin{subfigure}[b]{\textwidth}
%             \centering
%             \includegraphics[width=\textwidth]{figures/SeminarRoom_top_real.png}
%             \caption{}
%             \label{fig:schiavoni_top_real}
%         \end{subfigure}
%         %%
%         \begin{subfigure}[b]{\textwidth}
%             \centering
%             \includegraphics[width=\textwidth]{figures/SeminarRoom_top_VR.png}
%             \caption{}
%             \label{fig:schiavoni_top_VR}
%         \end{subfigure}
%     \end{minipage}
%     \caption{Plan view of the seminar room (a) and its \acrshort{3d} model in \acrshort{vr} (b)}
%     \label{fig:seminarroom_top}
% \end{figure}
% \begin{figure}[ht]
% \centering 
%     \begin{minipage}{7.8cm}
%     \centering
%         \begin{subfigure}[b]{\textwidth}
%             \centering
%             \includegraphics[width=\textwidth]{figures/SeminarRoom_real.png}
%             \caption{}
%             \label{fig:schiavoni_real}
%         \end{subfigure}
%         \begin{subfigure}[b]{\textwidth}
%             \centering
%             \includegraphics[width=\textwidth]{figures/SeminarRoom_VR.png}
%             \caption{}
%             \label{fig:schiavoni_VR}
%         \end{subfigure}
%     \end{minipage}
%     \caption{The real seminar room (a) and its \acrshort{3d} model in \acrshort{vr} (b)}
%     \label{fig:seminarroom}
% \end{figure}

\subsection{OSC protocol setup}
    The interaction between the \acrshort{vr} software and the external sound processing application is implemented through additional components taken from a Unity extension asset, \texttt{extOSC}\footnote{https://assetstore.unity.com/packages/tools/input-management/extosc-open-sound-control-72005}, that uses \acrshort{udp} to send customized \acrshort{osc} messages on an IP network.
These components are \texttt{C\#} scripts attached to all the \acrshort{ui} elements and to the main camera object, which represents the point of view of the user.
Each interaction with \acrshort{ui} elements triggers a different \acrshort{osc} message according to the specific function of the element.
In particular, buttons on the seat selection menu trigger a string type message that encodes the chosen seat position in order to recall the correct \acrshort{arir} in the sound processing phase.
Also, during the entire evaluation session, the \texttt{extOSC} components attached to the main camera, transmit two distinct \acrshort{osc} messages at frame rate, encoding the camera’s positional and rotational \acrshort{3d} coordinates.
%\begin{table}[]
%\begin{tabular}{|l|l|l|}
%\hline
%\textbf{\acrshort{ui} element} & \textbf{Data type} & %\textbf{Range} \\ \hline
%Toggle button       & boolean            & (0, 1) \\ \hline
%Slider              & integer            & (0, ..., 100) \\ \hline
%Seat button         & string             & \begin{tabular}[c]{@{}l@{}}(r1\_p1 … r1\_p5; \\ ... ; \\ r5\_p1 ... r5\_p5)\end{tabular} \\ \hline
%\end{tabular}
%\caption{\acrshort{osc} messages triggered by each \acrshort{ui} element.}
%\label{tab:osc_messages}
%\end{table}
\subsection{Interactive audio content and data management}
The second component of the proposed evaluation platform is a Max project running on a \acrshort{pc}.
It handles all the \acrshort{osc} messages received from the \acrshort{vr} system and streams spatial audio in real-time according to user inputs on the \acrshort{gui} and user positional and rotational data.
This component also takes care of collecting all the evaluation results and saving them in text files together with user behaviour data (time tracking, evaluated attribute and selected seat).
It also serves as a control panel for the test administrator, allowing him/her to view and change settings before and during tests.

\begin{figure*}[ht]
 \centerline{
 \includegraphics[width=\textwidth]{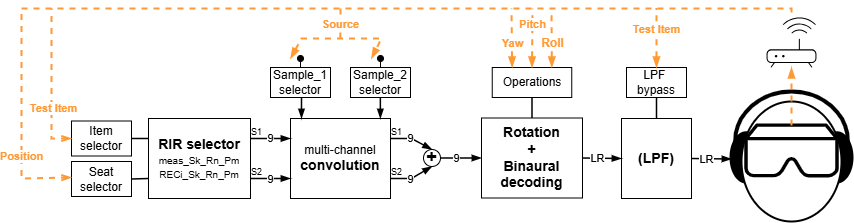}}
 \caption{Interactive spatial audio processing schematics.}
 \label{fig:spatializer_scheme}
 \vspace{-2mm}
\end{figure*}

The spatial audio processing chain is reported in \figref{fig:spatializer_scheme}.
All the measured and reconstructed \acrshort{arir}s are imported in a dedicated buffer and recalled by the \acrshort{rir} selector whenever the assessor selects a spot from the seat selection menu (\figref{fig:seatselection_VR}) or switches between test items on the evaluation menu (\figref{fig:MUSHRA_VR}).
Test administrators can select or import anechoic mono audio samples for each sound source.
These are separately convolved with the related \acrshort{arir}s by the \texttt{multiconvolve$\sim$} object from HISSTools package \cite{harker2012hisstools} before being summed up.
The resulting multichannel audio signal is sent to the SPARTA ambiBIN plugin \cite{mccormack2019sparta} that operates real-time rotation and binaural decoding.
The output of the whole pipeline is a binaural audio stream that is sent to a pair of closed-ear headphones.
An optional low-pass filter can be switched on if needed, e.g. to filter the reference signal and use it as an anchor in \acrshort{mushra}\cite{series2014method} tests.

% \begin{figure}[!h]
%  \centerline{
%  \includegraphics[width=6.8cm]{figures/VR-PTOLEMAIC_spatializer.png}}
%  \caption{Interactive spatial audio processing schematics.}
%  \label{fig:spatializer_scheme}
% \end{figure}

\section{System validation}\label{sec:eval_tests}

    To validate the proposed VR-based evaluation framework, we used \acrshort{sfr} as a test scenario and conducted listening tests within the VR environment.
\acrshort{sfr} is a fundamental task in spatial audio, focused on estimating the pressure field in regions where direct measurements cannot be obtained.
Capturing the acoustic field over a broad spatial area and across the entire audible frequency range requires extensive experimental resources, which is often not feasible.
Instead, \acrshort{sfr} techniques employ a limited set of observations to interpolate or extrapolate the acoustic field, offering a more practical alternative \cite{damiano_zero-shot_2024}.
%To evaluate the effectiveness of our proposed system for \acrshort{sfr} assessment, we conducted listening tests within a VR environment.
Using a \acrshort{mushra} test methodology, participants assessed various \acrshort{sfr} methods, comparing artificially generated signals against reference signals taken from \cite{miotello_homula-rir_2024}.
%Implemented on a Meta Quest 2 headset, 
%The system enabled controlled evaluation of sound quality and spatial perception.
The collected responses, along with participant reliability metrics, were analyzed to assess both the accuracy of the reconstruction methods and, more importantly, the suitability of the VR-based evaluation framework in the spatial audio context.

\subsection{Evaluation procedure}
%Listening tests were performed using a \acrshort{mushra}-like test methodology.
%The test procedure is composed of two phases.
During the procedure, participants were first asked to complete a familiarization phase to become acquainted with the tasks, the audio samples, the controllers, the teleportation system and the \acrshort{ui} used to evaluate reconstruction algorithms. This was followed by the assessment phase, consisting in three evaluation trials.
In each trial, a different audio sample is convolved with the $25$ \acrshortpl{arir} generated using 4 \acrshort{sfr} algorithms that the assessor has to evaluate according to 4 attributes with respect to an explicit reference signal.
The anechoic mono audio samples (\numrange{10}{15} \unit{\s} in length) were chosen from two different datasets.
Specifically, the test excerpts were taken from a female voice sample and a male voice sample in the EARS dataset \cite{richter_ears_2024}, along with an instrumental performance from the URMP dataset \cite{li_creating_2019}.

%\begin{itemize}
%    \item Female voice: \textit{p006/freeform\_speech 01.wav}
%    \item Male voice: \textit{p001/emo\_neutral\_sentences.wav}
%    \item Instrumental sound: %\textit{AuSep\_1\_vn\_02\_Sonata.wav}
%\end{itemize}
%Anechoic mono audio samples are selected from two different datasets,\tabref{tab:test_samples} for details.
For the evaluation of each attribute, five different audio stimuli were proposed. Besides the explicit reference, which is the audio stream of the sample convolved with the measured \acrshort{arir}s taken from the HOMULA-RIR dataset\cite{miotello_homula-rir_2024}, the stimuli to be evaluated were:
\begin{itemize}[leftmargin=*]
    \item Hidden reference $S_{\mathrm{ref}}$ - Sample convolved with the measured \acrshortpl{arir} presented without explicit identification to determine evaluators reliability;
    \item Non-parametric solution $S_A$ - Sound-field reconstructed employing an amplitude matching method \cite{abe2022amplitude};
    \item Low audio quality solution $S_{lp}$ - Low-pass filtered version of the reference with a cutoff frequency of $\SI{3.5}{\kilo\hertz}$, which reduces frequency response quality while preserving spatial response quality;
    \item Parametric solution $S_1$ - Sound-field reconstructed employing a virtual miking parametric methodology \cite{pezzoli_parametric_2020}.
\end{itemize}
Attributes, each one characterized by a specific values range, are chosen in order to collect ratings on both sound quality and spatial audio quality:
\begin{description}[style=unboxed,leftmargin=0cm]
   \item[Basic Audio Quality] [0 - 100] - Global attribute used to judge all detected differences between the reference and the object \cite{series2014method}.
   \item[Localizability] [More difficult - Easier] - Attribute that correlates with the perceived spatial extent of a source. Localizability \cite{lindau2014spatial} is low when spatial extent and location of a sound source are difficult to estimate or appear diffuse. It is high if a sound source is clearly delimited.% Low/high localizability is often associated with high/low perceived extent of a sound source.
   \item[Spatial Quality] [Low Quality - High Quality] - A measure of the ability of the item to acoustically describe the presented scene with respect to the reference.
    Takes into account all spatial characteristics, e.g., depth, width, spatial distribution, reverberation, spatialization, distance, envelopment, immersion.
   \item[Timbral Quality] [Low Quality - High Quality] - How accurately the item maintains the original harmonic content, tone color, and spectral balance of the sound with respect to the reference. 
\end{description}
Test sessions were carried out on a Meta Quest 2 headset with Sennheiser HD380 Pro headphones and a Behringer MicroAMP HA400 headphone amplifier while test administration and audio processing were carried out on a laptop equipped with an Intel Core i7-7700HQ CPU and 16GB of RAM.

\section{Results and discussion}\label{sec:discussion}

\begin{figure}[ht]
    \centerline{
    \includegraphics[width=7.8cm]{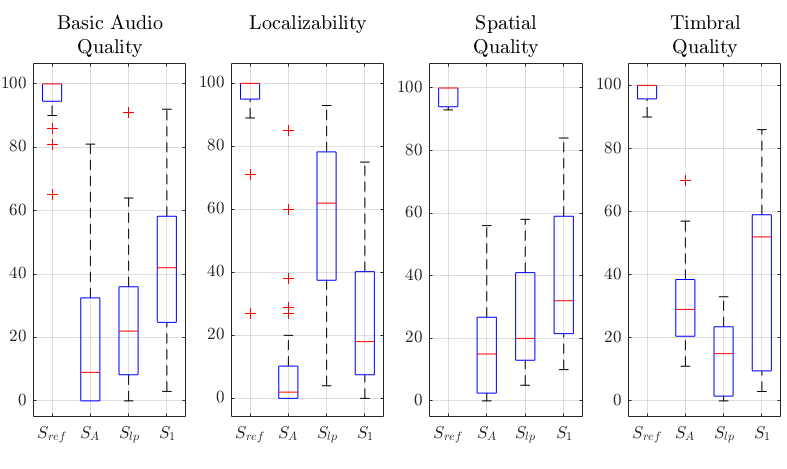}}
    \caption{Results of the evaluations aggregated for attribute.}
    \label{fig:results}
    %\vspace{-2mm}
\end{figure}

The proposed VR-based system was evaluated considering its ability to facilitate subjective spatial audio assessment and its impact on user interaction. 
Fifteen participants (13 male, 2 female) took part in the listening tests, with an average age of $26.9$ years ($\operatorname{SD} = 4.0$). 
All but one held a university degree, and all had prior experience in music. None reported hearing impairments, while three participants had no previous experience with \acrshort{vr}.
%All but one were students enrolled in the Music and Acoustic Engineering MSc program or employees at Politecnico di Milano. None reported any hearing impairments, and three had no prior experience with \acrshort{vr}.
Following the guidelines outlined in \cite{series2014method}, four assessors were excluded from the aggregated analysis, as they rated the hidden reference condition below a specific threshold for more than $15\%$ of the test items.
The \acrshort{mushra} test results are summarized in \figref{fig:results}, organized by the evaluated attributes.
In particular, the results obtained show that participants were able to distinguish between the different reconstruction methods with a high degree of consistency.
Although not central to this study's validation, these findings highlight the practical efficiency, usability and utility of the proposed evaluation framework in gathering test results for spatial audio applications.

\begin{figure}[ht]
    \centerline{
    \includegraphics[width=6.5cm, height=4.5cm]{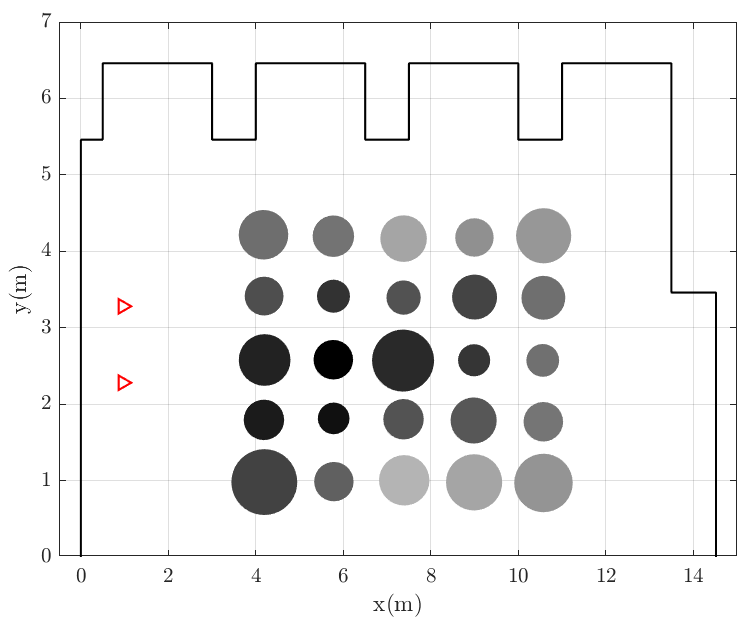}}
    \caption{User behaviour tracking. Circles dimension represents the time spent in that position. The gray scale represents the number of teleportation occurrences.}
    \label{fig:results_ubt}
    \vspace{-2mm}
\end{figure}

Additionally, user behaviour within the virtual scene was analyzed through head-tracking and teleportation data to gain insights into how participants interacted with the spatial environment during the evaluation. \figref{fig:results_ubt} presents the teleportation tracking results. In the figure, each position where a participant teleported is represented by a circle. The diameter of each circle corresponds to the amount of time spent in that location, providing an indication of where users preferred to linger during the listening tasks. Meanwhile, the greyscale shading of the circles reflects the frequency of teleportation events to that specific position: the darker the shade, the more often users moved to that spot.
Participants predominantly teleported in frontal positions, likely because these areas provided clearer spatial cues and more informative perspectives for evaluating source localization and sound field characteristics. Nonetheless, they also explored the other available positions throughout the room with relative consistency.

In addition to behavioural tracking, participants completed a questionnaire at the end of the session to assess user experience and usability. Most users described the system as intuitive in terms of user interface and overall user experience, with several highlighting that the VR interface helped them better perceive differences in spatial attributes. 
A few users reported mild discomfort associated with wearing a head-mounted display for extended periods, primarily due to the physical weight of the device during prolonged use.
These results support the system’s potential as an effective platform for spatial audio evaluation, combining perceptual assessment with behavioural data and providing an informative testing environment.

% \begin{figure}[ht]
%     \centering
%     \includegraphics[width=6.8cm, height=4cm]{figures/quest_boxplots.eps}
%     \caption{Results of the questionnaire about the comfort of the evaluation experience. Q1: time duration; Q2: wearing an \acrshort{hmd}; Q3: wearing headphones; Q4: use of VR controllers; Q5: \acrshort{ui}; Q6: \acrshort{ux}.}
%     \label{fig:equest_boxplots}
% \end{figure}

\section{Conclusion and future work}
We introduced a VR-based system for evaluating spatial audio reproduction, integrating real acoustic measurements into an immersive and interactive environment. The system was assessed through structured listening tests, which included a comparative evaluation of \acrshort{sfr} algorithms, and focused on perceptual attributes such as audio quality, localizability, and spatial and timbral quality. Results demonstrated that the platform effectively supports the evaluation of spatial audio algorithms, offering perceptual feedback comparable to traditional setups. Future work will refine the user interface and tracking analysis, with a focus on integrating detailed behavioural metrics. The simulation framework is planned for release as an open research tool.

\section{Acknowledgments}
\thanks{This work has been funded by "REPERTORIUM project. Grant agreement number 101095065. Horizon Europe. Cluster II. Culture, Creativity and Inclusive Society. Call HORIZON-CL2-2022-HERITAGE-01-02.".}
%\thanks{This work was supported by the Italian Ministry of University and Research (MUR) under the National Recovery and Resilience Plan (NRRP), and by the European Union (EU) under the NextGenerationEU project.}
\thanks{This work was partially supported by the European Union -- Next Generation EU under the Italian National Recovery and Resilience Plan (NRRP), Mission 4, Component 2, Investment 1.3, CUP D43C22003080001, partnership on ``Telecommunications of the Future’' (PE00000001 -- program ``RESTART’'}

%\begin{table}[!h]
% \caption{Values of the main room acoustics parameters measured and simulated, and differences in JND’s..}
% \begin{center}
% \begin{tabular}{|l|l|}
%  \hline
%    Frequency (Hz) & 63 \\
%  \hline
%  Value1  & \conferenceyear\ \\
% \hline
% \end{tabular}
%\end{center}
% \label{tab:example}
%\end{table}

% For bibtex users:
%\bibliographystyle{jabbrv_unsrt}
%\bibliographystyle{jabbrv_ieeetr}
\bibliography{fa2025_template}

% For non bibtex users:
%\begin{thebibliography}{citations}
%\bibitem{Author:00}
%E.~Author.
%\newblock The title of the conference paper.
%\newblock In {\em Proc.\ of the European Society on Vibration
%  }, pages 000--111, Chania, Greece, 2018.
%
%\bibitem{Someone:10}
%A.~Someone, B.~Someone, and C.~Someone.
%\newblock The title of the journal paper.
%\newblock {\em Acta Acust united Ac}, A(B):111--222, 2010.
%
%\bibitem{Someone:04}
%X.~Someone and Y.~Someone.
%\newblock {\em The Title of the Book}.
%\newblock S. Hirzel, Stuttgart, Germany, 2012.
%
%\end{thebibliography}

\end{document}